\documentclass[12pt,preprint]{aastex}

\shorttitle{Slow Heating Model of Gamma-Ray Burst}
\shortauthors{Asano \& Terasawa}

\begin{document}

\title{
Slow Heating Model of Gamma-Ray Burst:\\
Photon Spectrum and Delayed Emission
}
\author{\scshape Katsuaki Asano and Toshio Terasawa}
\affil{Interactive Research Center of Science, Graduate School of Science,
Tokyo Institute of Technology, 2-12-1 Ookayama, Meguro-ku, Tokyo 152-8550, Japan}
\email{asano@phys.titech.ac.jp, terasawa@phys.titech.ac.jp}


\begin{abstract}
We propose a new
mechanism for the prompt emission of gamma-ray burst.
In our model electrons are continuously accelerated in the post shock region
via plasma turbulence.
Using the Monte Carlo technique, we mimic the second-order Fermi
acceleration due to plasma turbulence and obtain photon spectra.
Since the acceleration balances with the synchrotron cooling,
the observed low-energy spectral index is naturally explained.
The resultant spectra can be consistent with observed spectra
at least below $\sim 1$ MeV.
The model also predicts delayed GeV-TeV emission due to inverse Compton
and broad pulse profile of optical emission in some cases.
Although nontrivial assumptions are required
to reproduce MeV-GeV power-law spectra, the model implies the possibility
to explain various kinds of luminosity correlations.
\end{abstract}


\section{Introduction}
\label{sec:intro}

In the widely discussed internal shock scenario \citep[see, e.g., reviews by][]{pir05,mes06},
the prompt emission of gamma-ray bursts (GRBs) is due to
collisions among inhomogeneities within ultrarelativistic outflows,
which lead to formation of shocks.
The nonthermal photons, whose typical energy $\varepsilon_{\rm p}
\sim$ a few hundred keV,
are emitted from shock-accelerated electrons in highly magnetized plasma.
However, several open problems for the internal shock model 
have been pointed out such as the radiative efficiency,
various kinds of luminosity correlations, and so on.
In this paper, we focus on the two open problems:
the energy transfer problem and the low-energy spectral index problem.
The standard model postulates that a large fraction of the kinetic energy
carried by protons should be efficiently 
converted into that of relativistic electrons.
However, it is apparent that the Coulomb interaction cannot transport the internal 
energy of heated protons 
into electrons to achieve energy equipartition, because
the timescale of the Coulomb interaction is much longer
than the dynamical timescale.
While the simple first-order Fermi acceleration at the shock front
is assumed to transfer the energy into electrons in the standard model,
some unknown plasma processes may play an important role in the energy transfer.

The second problem is in the spectral shape of the prompt emission.
The observed spectra of GRBs are well fitted with the conventional
Band function \citep{ban93}; the photon number spectrum
$\propto \varepsilon^{\alpha} \exp{[-(2+\alpha)\varepsilon/\varepsilon_{\rm p}]}$
below $(\alpha-\beta)\varepsilon_{\rm p}/(2+\alpha)$,
and $\propto \varepsilon^{\beta}$ above it.
The typical fitted value of the low energy spectral index
is $\alpha=-1.0$ \citep{pre00}, while the standard model predicts
that photons from cooled electrons dominate the low energy region
below $\varepsilon_{\rm p}$, which leads to $\alpha=-1.5$.
To resolve this problem several alternative models, 
such as the thermal emission from photosphere 
\citep[and references therein]{mes00, iok07}, 
have been considered.
The Klein-Nishina effect on synchrotron self-Compton (SSC)
process, which can affect the
low energy synchrotron spectrum, has been discussed frequently
\citep{der01,bos09,nak09,wan09}.
Recently, \citet{pee06b} have suggested that the decay of magnetic fields
(\S \ref{sec:dec})
may resolve the problem in the low energy spectral index.
However, 
in that model, we have not yet found a reason why
the decay timescale should always be comparable to the cooling timescale
(\S \ref{sec:model}).
In a particle-in-cell (PIC) simulation for electron-positron plasma
\citep{cha08}, the decay timescale is close to the requirement of \citet{pee06b},
but long-term evolution of magnetic fields is still controversial,
partially because the effect of accelerated particles
is not quantitatively unveiled yet \citep{kes09}.

Motivated by these problems,
we propose an alternative model for the prompt emission of GRBs 
(\S \ref{sec:simple}).
In our model electron-heating (second-order Fermi acceleration)
due to plasma turbulence continues during photon emission in shocked plasmas
\citep{ghi99},
so that the resultant spectral index of low energy photons
can be consistent with the observations at least below $\sim 1$ MeV.
Since the assumed heating timescale
is longer than that in the standard scenario,
we call this model the ``slow heating model''.
After the free energy for the plasma instabilities is dissipated,
the magnetic fields may decay and the synchrotron emission will cease.
The further possibility to reproduce MeV-GeV power-law spectra 
within the framework of this model
is discussed in \S \ref{sec:var}.
In addition, our model naturally predicts delayed GeV-TeV emission due to inverse Compton
(IC) and broad pulse profile of optical emission under certain conditions (\S \ref{sec:del}).
Finally, we summarize the results of our model, and compare them with 
the observed luminosity correlations in GRBs (\S \ref{sec:sum}).
We should note that our synchrotron model is different from
the IC models using the same terminology ``slow heating'' in
\citet{pee06} and \citet{gia08},
in which the electrons are heated slowly
in a timescale comparable to the shell-expansion timescale
\citep{ghi99,ste04,vur09}.

\section{Magnetic Field: Generation and Decay}
\label{sec:dec}
In the standard scenario, the magnetic fields
are assumed to be generated/amplified in the region around shocks
via plasma instabilities, such as
the Weibel instability
\citep[e.g.,][]{kaz98,med99,sil03,nis05,kat07} 
or the two-stream instability due to high-energy particles
accelerated at the shock \citep[e.g.,][]{bel04}.
The difference in temperatures of electrons and protons may also arouse
some plasma instabilities.
Generated magnetic fields efficiently interact with
particles, whose Larmor radii are comparable to the typical scale
of the turbulence.
This situation is definitely different from the ideal magnetohydrodynamic (MHD)
approximation.
For example, 
where the electron Larmor radii are finite,
the off-diagonal terms in the electron pressure tensor
could appear and catalyze magnetic reconnection
\citep[e.g.,][]{moz09}.
The energy of magnetosonic perturbations can be
transferred to resonant particles via the transit-time damping process
\citep[e.g.,][]{sch98}.
It is natural, therefore,
to consider that the generated magnetic fields 
may decay via interaction with particles after the free energy
for instability excitation 
(anisotropy, inhomogeneity, different temperatures of
electrons and ions etc.) is dissipated
with a decay timescale $t_{\rm dec}$.
Actually, recent PIC simulations of electron-positron plasmas show that
magnetic turbulences induced by the Weibel instability
decay \citep[e.g.,][]{cha08,kes09}.
For $t >t_{\rm dec}$,
electrons stop emitting photons via synchrotron radiation.
Therefore,
the decay of magnetic fields may suppress the photon emission
from cooled electrons, which resolves the problem in the index $\alpha$.

\section{Numerical Model: Standard Case}
\label{sec:model}
\begin{figure}[h]
\centering
\epsscale{1.0}
\plotone{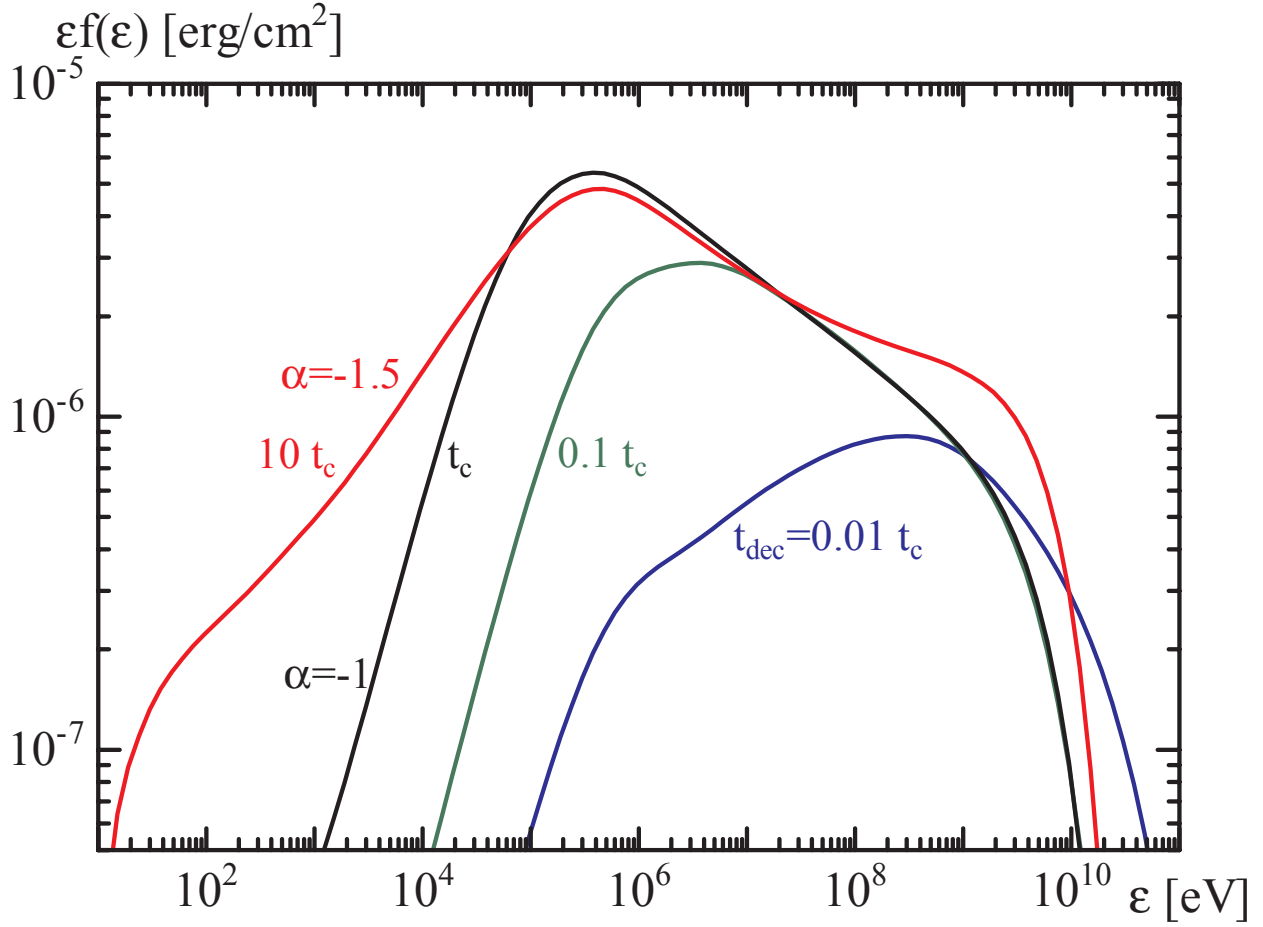}
\caption{Photon spectra for the standard model
with decaying magnetic fields varying the decay timescale
$t_{\rm dec}$. The assumed parameters are $p=2.5$, $E_{\rm sh}=10^{51}$ erg,
$\Gamma=300$, $R=2.7 \times 10^{14}$ cm ($\Delta t=0.1$ s),
$B=3200$ G, and $\gamma_{\rm e,m}=3900$.
\label{fig:test}}
\end{figure}

First let us revisit the effects of decay of magnetic fields
with the standard manner of the electron injection.
In Figure \ref{fig:test}, changing the decay timescale,
we show GRB spectra obtained by numerical calculations with the same code
in \citet[][details will be explained in the following section]{asa07}.
Throughout this paper, all spectra are shown in terms of the observed fluence versus
photon energy, assuming a GRB redshift of $z=0.1$.
The vertical axes denote $\varepsilon f(\varepsilon)$,
so that photon spectra with a spectral index $\alpha$
are plotted as $ \propto \varepsilon^{\alpha+2}$.
The model parameters are estimated as follows.
The emitting region for a pulse is 
a homogeneous shell expanding with the Lorentz factor
$\Gamma$ at radius $R$ from the central engine.
We adopt $l=R/\Gamma$ for the comoving width of the shell,
so that the pulse timescale in the observer frame is $\Delta t=R/\Gamma^2 c$
\citep{sar97}.
Here, we choose parameters,
$\Gamma=300$, $\Delta t=0.1$ s, which implies $R=2.7 \times 10^{14}$ cm.
 The energy density of accelerated electrons in the shell
$U_{\rm e}=\epsilon_{\rm e} U$
($U$ is the total energy density of the shocked plasma)
is a parameter that can be directly related to
the isotropic-equivalent energy of photons from a single pulse $E_{\rm sh}$
(here we adopt $10^{51}$ erg)
as $E_{\rm sh} = U_{\rm e}{\cal V}$,
where ${\cal V} \equiv 4 \pi R^3/\Gamma$ is the comoving volume.

In the standard scenario,
relativistic electrons are injected at the shock front
with a power-law energy distribution
$\dot{N}(\gamma_{\rm e}) \propto \gamma_{\rm e}^{-p}$ 
for $\gamma_{\rm e} \ge \gamma_{\rm e,m}$,
where $\gamma_{\rm e}$ is the electron Lorentz factor in the plasma rest frame.
This scenario requires a sharp low-energy cutoff for the electron injection spectrum;
the minimum Lorentz factor $\gamma_{\rm e,m}$ 
is evaluated in the literature by giving the energy density of electrons 
$U_{\rm e}=\epsilon_{\rm e} U$
together with the total number density of electrons.
Therefore, in the standard scenario,
$\gamma_{\rm e,m}$ has been conventionally described
by the phenomenological parameter $\epsilon_{\rm e}$,
though the energy scale corresponding to $\gamma_{\rm e,m}$ should be
derived from physics in relativistic plasmas.
Here, instead of $\epsilon_{\rm e}$,
we take $\gamma_{\rm e,m}$ to be a parameter,
because we do not concern the non-observable parameter $U$.

The photon energy $\varepsilon_{\rm p}$
 corresponding to $\gamma_{\rm e,m}$ is given by
\begin{eqnarray}
\varepsilon_{\rm p} \simeq
\frac{\hbar e B \gamma_{\rm e,m}^2}{m_{\rm e} c} \Gamma.
\end{eqnarray}
The cooling timescale for electrons of $\gamma_{\rm e,m}$ is written as
\begin{eqnarray}
t_{\rm c}(\gamma_{\rm e,m})=\frac{6 \pi m_{\rm e} c}{\sigma_{\rm T} B^2 \gamma_{\rm e,m}},
\label{tc}
\end{eqnarray}
where $\sigma_{\rm T}$ is the Thomson cross section.
With a non-dimensional parameter $\epsilon_B$, 
the magnetic energy density $U_B \equiv B^2/8 \pi$
is given as $\epsilon_B U= (\epsilon_B/\epsilon_{\rm e}) U_e$.
In Figure \ref{fig:test},
we set $B=$ 3200 G and 
$\gamma_{\rm e,m}=3900$, 
which correspond to
$\epsilon_B/\epsilon_{\rm e}=0.1$ 
and $\varepsilon_{\rm p} \sim$ a few hundred keV, respectively.

We numerically follow electron cooling via
synchrotron and IC emissions,
adopting the Klein-Nishina cross section,
and artificially stop the calculation
after $t_{\rm dec}$ to mimic the decay of magnetic fields.
The effects of
$\gamma \gamma$ pair production and synchrotron self-absorption
are also taken into account.
Since $\gamma_{\rm e,m} \varepsilon_{\rm p}/\Gamma > m_{\rm e} c^2$ in our choice,
the Klein-Nishina effect cannot be neglected for IC emission.
The dynamical timescale $t_{\rm dyn}=l/c=30$ s is much longer than 
the cooling time $t_{\rm c}=0.02$ s.

The blue curve with $t_{\rm dec} = 0.01 t_{\rm c}\ll t_{\rm c}$
corresponds to the slow cooling case \citep{sar98}.
In this case, $\varepsilon_{\rm p}$ is determined by
the lowest energy of electrons that can cool within the timescale
$t_{\rm dec}$. As is well known,
the index in the slow cooling case is
$\alpha=-(p+1)/2$, which is softer than the typical observed $\alpha$
for our choice of $p=2.5$.
If we adopt a very hard injection index
$p \simeq 1$, $\alpha$ can be $\sim -1$, but
the high-energy index $\beta = -(p+2)/2 \sim -1.5$ contradicts
the typical value $\beta < -2$.
On the other hand, 
for the two cases, $t_{\rm dec} = 0.1 t_{\rm c}$ and $ t_{\rm c}$ 
(green and black curves),
the index becomes $\alpha \simeq -1$ below $\varepsilon_{\rm p}$.
These cases are what \citet{pee06b} suggested
to solve the problem of the low-energy spectral index.
 
It is further seen in Figure \ref{fig:test} that
for the case of $t_{\rm dec} = 10 t_{\rm c}$ (red curve)
the spectrum shows $\alpha \simeq -1.5$.
This is the prediction by the standard model
as referred in the introduction:
electrons injected with $\gamma_{\rm e}=\gamma_{\rm e,m}$
are cooled after $t=t_{\rm c}$, and the low-energy spectrum
becomes soft owing to emissions from such cooled electrons.
It is noted that this case shows
a spectral bump in the GeV band due to IC emission,
whose contribution is boosted up
by enhancement of low-energy seed photons.
We can also see cutoffs above 10 GeV and below 30 eV.
They are $\gamma \gamma$-absorption and synchrotron self-absorption, respectively.

The above results indicate that only with the case,
$t_{\rm dec} \sim t_{\rm c}(\gamma_{\rm e,m})$,
the decaying magnetic field can explain the low-energy index $\alpha$.
However, there is no definite physical reason to expect such a
matching between $t_{\rm dec} $ and $ t_{\rm c}(\gamma_{\rm e,m})$.
The jitter radiation \citep{med00,fle06}
instead of the synchrotron radiation is worthwhile to consider,
because the typical scale of turbulence excited by plasma instabilities
can be much shorter than the Larmor radii of radiating electrons.
While the typical photon energy in the jitter radiation,
which is determined by the coherence scale of the disturbed magnetic field,
differs from in the usual synchrotron radiation,
the introduction of the jitter radiation
does not significantly change the low-energy spectral shape:
$\alpha$ remains 
$\simeq -1.5$ as long as $t_{\rm dec} \gg t_{\rm c}$ (fast cooling).

\section{Slow Heating Model: start}
\label{sec:simple}

\begin{figure}[h]
\centering
\epsscale{1.0}
\plotone{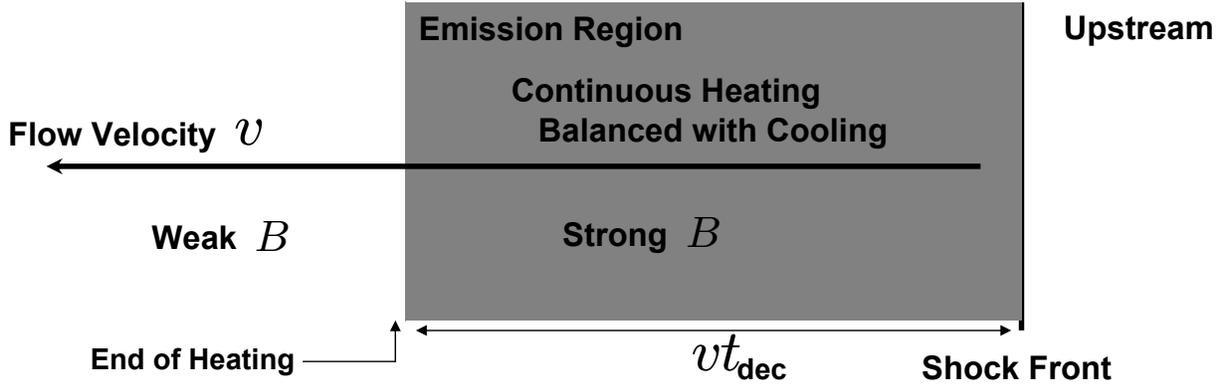}
\caption{Schematic pictures of the standard model and
slow heating model. In the standard picture, radiation
from the area behind the emission region cannot be neglected,
as long as the magnetic field remains.
On the other hand, the end of heating due to plasma turbulence
and decay of the magnetic field are naturally simultaneous in the slow heating model.
\label{fig:sch}}
\end{figure}

As we mentioned in \S \ref{sec:dec}, turbulent magnetic fields
may be generated in the plasmas around shocks.
Such turbulent waves may play a role in energy transfer from protons
to electrons until the magnetic fields decay.
In this section we present our new model, the slow heating model,
to resolve the index problem.
While the standard picture postulates a prompt acceleration of electrons,
whose timescale is much shorter than $t_{\rm c}(\gamma_{\rm e,m})$,
our model assumes slower energy transfer from the background plasma to
electrons via some unknown plasma instabilities (see Figure \ref{fig:sch}).

In order to mimic the energy transfer we consider
the second-order Fermi acceleration, even though there may exist
not only Alfv\'en waves but also other types of acceleration mechanisms
such as electric fields around ion current channels \citep{hed04}
or coherent wave-particle interactions resulting from parametric instabilities
\citep{mat09},
which can also contribute to particle acceleration.
When a particle is scattered by a wave or magnetized cloud
preserving its energy in the wave (cloud) frame,
twice Lorentz transformations give us energy gain due to this collision
$\xi \equiv \Delta E/E$ as
\begin{eqnarray}
\xi =\gamma_0^2 \left(
1-\beta_0 \mu_1+\beta_0 \mu'_2-\beta_0^2 \mu_1 \mu'_2
\right)-1,
\end{eqnarray}
where $\gamma_0=1/\sqrt{1-\beta_0^2}$, $\mu_1$, and $\mu'_2$ are
the Lorentz factor of the wave (cloud), cosines of incident angle
in the reference frame, and scattering angle in the wave (cloud) frame,
respectively.
If the wave velocity is non-relativistic ($\beta_0 \ll 1$),
the mean energy gain $\overline{\xi} \sim \beta_0^2$
under the assumption of isotropic wave distribution $\overline{\mu_1}=-\beta_0/2$
and isotropic scattering $\overline{\mu'_2}=0$.
So the second-order Fermi acceleration is a slower acceleration process
than the first-order one ($\overline{\xi} \sim \beta_0$) in non-relativistic cases.
However, GRB internal shock is relativistic so that we can expect
turbulent magnetic fields with $\beta_0 \sim 1$.

The Fokker-Planck equation for ultrarelativistic particles can be written as
\begin{eqnarray}
\frac{\partial N}{\partial t}=\frac{\partial}{\partial E}
D_{EE} \frac{\partial N}{\partial E}-\frac{\partial}{\partial E}
\left[ \left(2 \frac{D_{EE}}{E}-\dot{E}_{\rm cool} \right) N \right],
\label{FP}
\end{eqnarray}
where $D_{EE}$ is the energy diffusion coefficient \citep[see, e.g.,][]{liu06}.
Defining the mean free time of particles $t_{\rm coll}$,
we can write
\begin{eqnarray}
D_{EE}=\frac{\overline{\xi} E^2}{2 t_{\rm coll}},
\end{eqnarray}
and the acceleration timescale $t_{\rm acc}=t_{\rm coll}/\bar{\xi}$.
At present we have no reliable model of relativistic turbulence in GRBs.
For reference, let us look in stochastic acceleration in non-relativistic plasma.
When we express the diffusion coefficient as $D_{EE} \propto E^n$
($t_{\rm acc} \propto E^{2-n}$), we obtain $n=m$
for isotropic Alfv\'en turbulence of spectral energy density per unit wavenumber
$W_k \propto k^{-m}$ \citep[see, e.g.,][]{mil89}.
The model with $n=2$ is often adopted
for small scale MHD turbulences \citep[see, e.g.,][]{liu06}.
Another value $n=5/3$ is frequently used for the Kolmogorov turbulence.
The strong turbulence limit (Bohm limit), where the mean free path
becomes comparable to Larmor radii, corresponds to the case of $n=1$,
where the dependence on $m$ disappears.
Relativistic shocks in electron-positron plasmas in PIC simulations
\citep{cha08} generate magnetic turbulence with $m=0$ for small $k$
($kc \ll$ plasma frequency), but $m \simeq 2$ for large $k$.
Although we have no definite shape of $D_{EE}$ for GRBs yet,
future long-term PIC simulations may reveal the property and evolution
of magnetic turbulence.

From eq. (\ref{FP}) we may write
$\overline{\Delta E^2}=(\overline{\xi}-\overline{\xi}^2) E^2$,
so that we assume the probability function of $\xi$ per collision as
a Gaussian form,
\begin{eqnarray}
P(\xi) =\frac{1}{\sqrt{2 \pi}\sigma} \exp{\left[
-\frac{(\xi-\overline{\xi})^2}{2 \sigma^2}\right]},
\quad \sigma=\sqrt{\overline{\xi}-\overline{\xi}^2}.
\label{prob}
\end{eqnarray}
Hereafter, we assume a constant value of $\overline{\xi}=0.1$.
Considering synchrotron and IC emissions,
the energy loss rate due to radiative cooling is expressed as
\begin{eqnarray}
\dot{E}_{\rm cool}=\frac{4}{3} \sigma_{\rm T} c \gamma_{\rm e}^2
U_B \left( 1+K(\gamma_{\rm e}) \frac{U_{\rm ph}}{U_B} \right),
\label{cool}
\end{eqnarray}
where $U_{\rm ph}$ and $K(\gamma_{\rm e})$ are the photon energy density
and the correction coefficient due to the Klein-Nishina effect, respectively.
In the Thomson limit, $K(\gamma_{\rm e})=1$.

We employ the Monte Carlo numerical code of \citet{asa07} to follow
the radiative cooling and stochastic energy gain/loss processes
according to eqs. (\ref{cool}) and (\ref{prob}) with a time step,
\begin{eqnarray}
\delta t=\min(t_{\rm coll}/30,E/\dot{E}_{\rm cool}/30,t_{\rm dec}/30),
\end{eqnarray}
and at $t=t_{\rm dec}$, we artificially halt the calculations to mimic
the decay of magnetic fields.
For each time step, we judge the occurrence of collision and estimate
energy loss due to radiation using random numbers.
If a collision occurs, the energy gain/loss due to the collision
is counted with evaluated $\xi$.

Since highly disturbed magnetic fields are assumed,
it is meaningful to consider the jitter radiation \citep{med00,fle06}.
But, for simplicity, we consider usual synchrotron radiation,
using the synchrotron function \citep{ryb79}.
As for IC emission we numerically estimate the spectral photon emission rate
and $K(\gamma_{\rm e})$ by integrating the photon energy distribution
given in advance
with the Klein-Nishina cross section $\sigma_{\rm KN}$ \citep{ryb79}.
We assume a uniform and isotropic photon field within a shell with width $l=R/\Gamma$
in the shell frame.
To obtain photon spectra,
the energy distributions of photons and particles are simulated iteratively
until the resultant spectrum and presupposed spectrum are identical.

In addition we take into account
$\gamma \gamma$ pair production and synchrotron self-absorption.
However, these processes are not so important in this paper,
so that we omit the explanation of the method to include these effects
\citep[see][]{asa07}.

\begin{figure}[h]
\centering
\epsscale{0.8}
\plotone{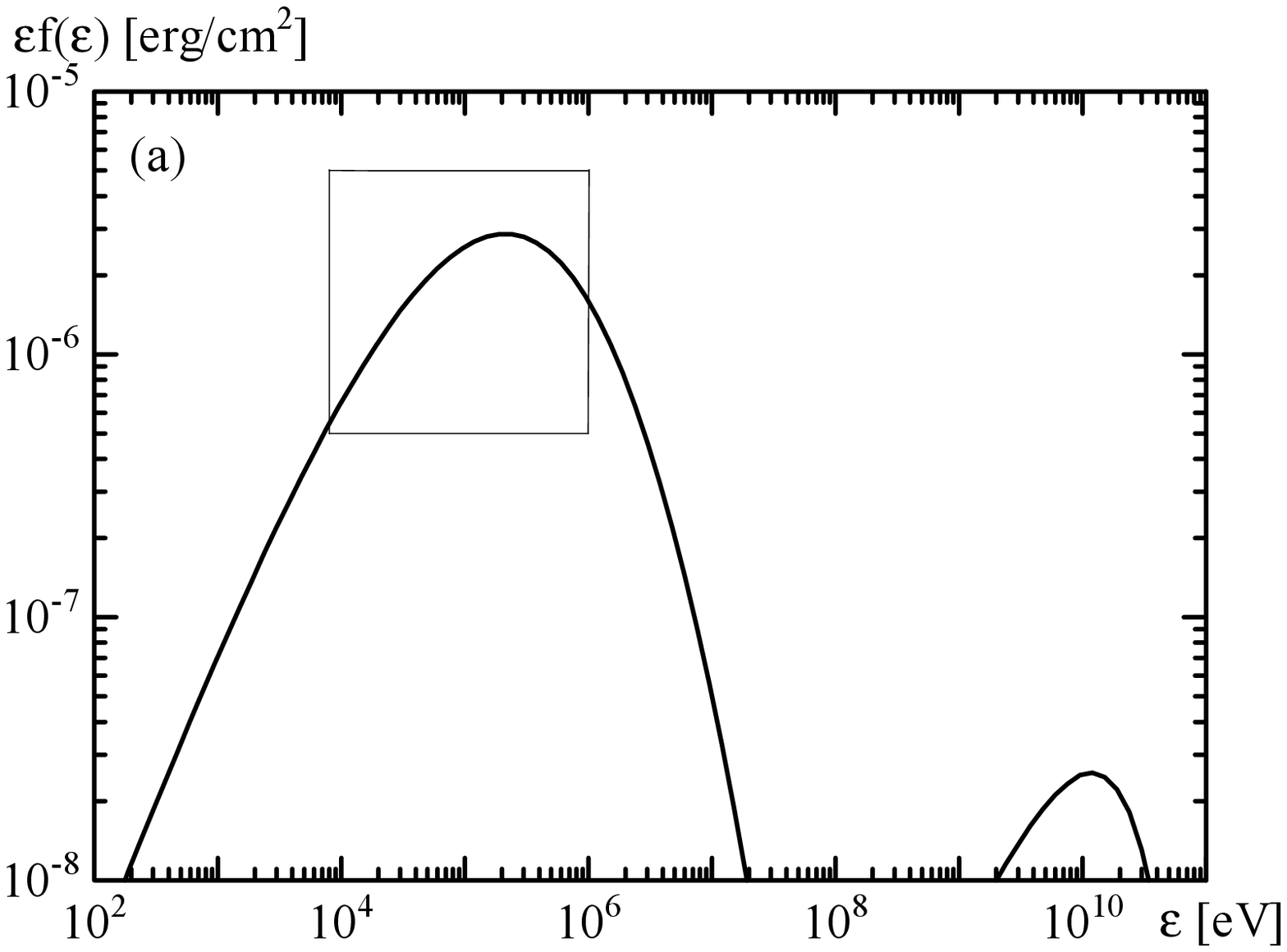}
\plotone{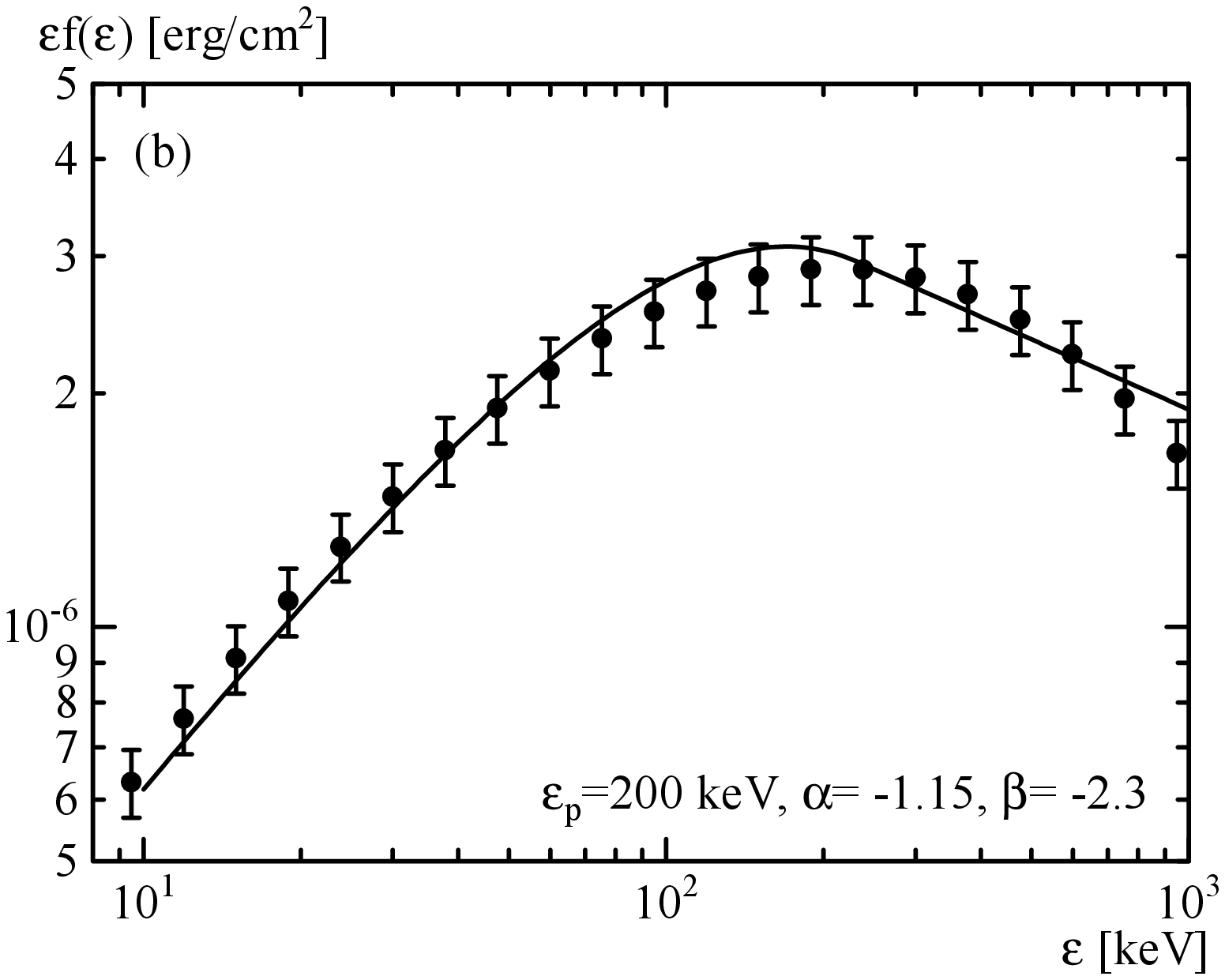}
\caption{
(a) Spectrum for the slow heating model with constant $\xi$ and $t_{\rm coll}$.
(b) Zoom up of the upper figure (the area enclosed by the rectangle)
for 10 keV -- 1 MeV range.
The data points are calculated spectra with 10 \% error putted in by hand,
and the solid line is the Band function with $\varepsilon_{\rm p}=200$ keV,
$\alpha=-1.15$, and $\beta=-2.3$.
\label{fig:2nd}}
\end{figure}

The studies for plasma turbulences in the post shock region by many authors
are ongoing now.
Although remarkable development is seen in
recent PIC simulations and MHD simulations \citep[e.g.,][]{wzha09},
a definite picture of shocked plasma is not understood yet.
Here, we consider a simple toy model assuming
that $D_{EE} \propto E^2$, which means
that $t_{\rm coll}$ does not depend on the energy of electrons.
Although we take into account IC emission,
it is not a main subject to discuss in this paper.
In order to concentrate on synchrotron photon spectra,
we adopt a stronger magnetic field $B=10^4$ G.
The other parameters are the same as those in \S \ref{sec:model}
except for the electron injection.
The typical photon energy is expected to be emitted from
electrons, whose energy loss rate is balanced with
the second-order Fermi acceleration.
Therefore, we adjust $t_{\rm coll}$ to make $t_{\rm acc}=t_{\rm c}$
at $\gamma_{\rm e}=\gamma_{\rm typ}=3100$
($t_{\rm c}(\gamma_{\rm typ}) \equiv t_{\rm c,typ} \sim 2 \times 10^{-3}$ s)
that implies the typical photon energy $\varepsilon_{\rm p}(\gamma_{\rm typ}) \sim$
a few hundred keV.
The Klein-Nishina effect is important for electrons of $\gamma_{\rm e}=\gamma_{\rm typ}$
even in this case.
The number of electrons is roughly
adjusted to make $E_{\rm sh}=10^{50}$-$10^{51}$ erg
considering the heating rate and $t_{\rm dec}$
(we may not exactly forecast the final photon energy in advance).
Below (above) $\gamma_{\rm typ}$ the acceleration timescale is shorter (longer) than
the synchrotron cooling timescale.
The heating due to turbulence reduces the effective number of
electrons below $\gamma_{\rm typ}$ so that the low-energy photon spectrum
is expected to be harder than the standard one (the red line in Figure \ref{fig:test})
even for $t_{\rm dec} \gg t_{\rm c,typ}$.
At $t=0$ electrons are injected with monochromatic energy
of $\gamma_{\rm e}=\gamma_{\rm inj}=\gamma_{\rm typ}/10$.
We have confirmed that a run with 5000 particle histories
is enough to converge.
In order to verify that the low-energy photon spectra
become hard enough ($\alpha \sim -1$) even for a longer decay timescale
than the cooling timescale,
we adopt $t_{\rm dec}=30 t_{\rm c,typ}$.
The result is shown in Figure \ref{fig:2nd}.

Photons at the spectral peak are emitted from electrons of
$\gamma_{\rm e} \sim \gamma_{\rm typ}$ as anticipated in advance.
Since electrons are accelerated immediately,
the injection parameter $\gamma_{\rm inj}$ does not affect the resultant spectrum very much.
The low-energy spectral index is well approximated as $\sim -1$.
The spectral bump at $\sim$ GeV is due to IC emission,
whose contribution is small because of high $B$
($\epsilon_B/\epsilon_{\rm e} \sim 3$) and Klein-Nishina effect.
Compared to Figure \ref{fig:test},
the lower density of target photons ($\geq 2$ MeV) weakens the $\gamma \gamma$
absorption effect on IC emission.
The overall shape of the spectrum is different from
the Band function.
We magnify the spectrum for 10 keV--1 MeV range in Figure \ref{fig:2nd}(b).
In this energy range, the spectrum with
artificial errors (10\%) does not contradict the Band function very much,
even though the model is quite simple.

\section{Slow Heating Model: modification}
\label{sec:var}

While the model spectra in \S \ref{sec:simple} may be fitted
with the Band function below $\sim$ MeV,
some GRBs show power-law spectra in the MeV-GeV range
with $\beta \sim -2$
\citep[see][as one of the recent examples]{abd09}.
As numerous simulations for the broadband prompt emission spectrum
have shown
\citep[e.g.,][]{pee06,gup07,asa07,bos09},
the usual simple power-law injection for electrons can easily reproduce
the MeV-GeV power-law spectra.
In the slow heating model, one of the simplest interpretation
to overcome this difficulty
is that such power-law spectra are superpositions
of multiple components with different $\varepsilon_{\rm p}$.
This explanation requires fine adjust of the amplitude of
multiple components.
Although this interpretation remains viable so far,
we search for alternative ideas within the slow heating model
in this section.

In \S \ref{sec:simple}, the index $n$ of energy dependence of
$D_{EE} (\propto E^n)$ was taken 2. 
To make the resultant energy spectrum harder,
we first test the cases with $n>2$
(note that $t_{\rm acc}<t_{\rm c}$ for $n>3$ in higher energy range).
Our simulations show that the spectral shape becomes close to a power-law function
as $n$ increases.
However, even for an extreme choice of $n=3$ ($t_{\rm acc} \propto E^{-1}$),
the resultant index $\beta=-3.2$ is still not hard enough.

\begin{figure}[h]
\centering
\epsscale{1.0}
\plotone{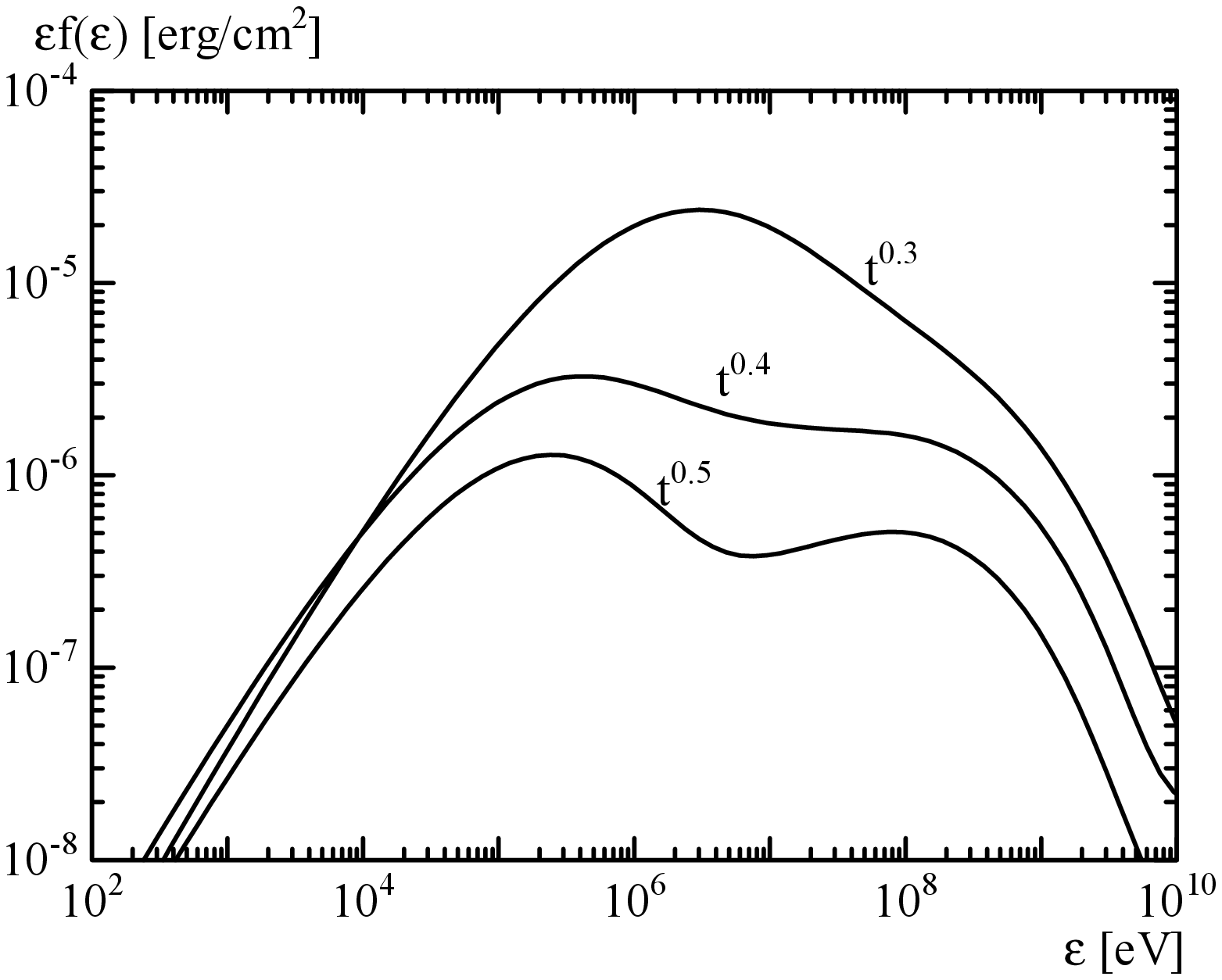}
\caption{
Spectra for the model of $t_{\rm coll} \propto t^{\chi}$
($D_{EE} \propto t^{-\chi}$,
$\chi=0.3, 0.4$, and $0.5$), $\gamma_{\rm inj}=10 \gamma_{\rm typ}$,
and $t_{\rm dec}=300 t_{\rm c,typ}$.
\label{fig:Tdep}}
\end{figure}

Since we consider the decay of the magnetic fields,
it is natural to let $D_{EE} = 2 \bar{\xi} E^2/ 2 t_{\rm coll}$ 
also time (or equivalently distance from the shock front) dependent.
Our toy model assumes a power-law shape as
$t_{\rm coll} \propto \gamma_{\rm e}^0 t^{\chi}$ 
with upper and lower limits. Namely,
\begin{eqnarray}
t_{\rm coll}=\min\left[\bar{\xi} t_{\rm c,typ},
\max\left\{ t_{\rm min} \left(
\frac{t}{t_{\rm min}} \right)^{\chi},
t_{\rm min} \right\}
\right],
\end{eqnarray}
where $t_{\rm min} \equiv \bar{\xi} t_{\rm c,typ}/100$ in our model.
In this case, the initial short timescale of acceleration makes
$\varepsilon_{\rm p}$ higher, and as the acceleration timescale
elongates
with time, $\varepsilon_{\rm p}$ will be settled
around a few hundred keV.
Here, to harden spectra, we adopt $\gamma_{\rm inj}=10 \gamma_{\rm typ}$.
Since the acceleration time from $\gamma_{\rm inj}$ to $100 \gamma_{\rm typ}$
is $\sim 10 t_{\rm min} \gg t_{\rm min}$,
the highest photon energy may be $\sim 100$ MeV emitted from
electrons of $\gamma_{\rm e}=\gamma_{\rm inj}$
(10 GeV photons from electrons of $\gamma_{\rm e}=100
\gamma_{\rm typ}$ may not be produced so much).
Assuming $t_{\rm dec}=300 t_{\rm c,typ}$ (other parameters are the same as before),
we calculate spectra (see Figure \ref{fig:Tdep}).

\begin{figure}[h]
\centering
\epsscale{1.0}
\plotone{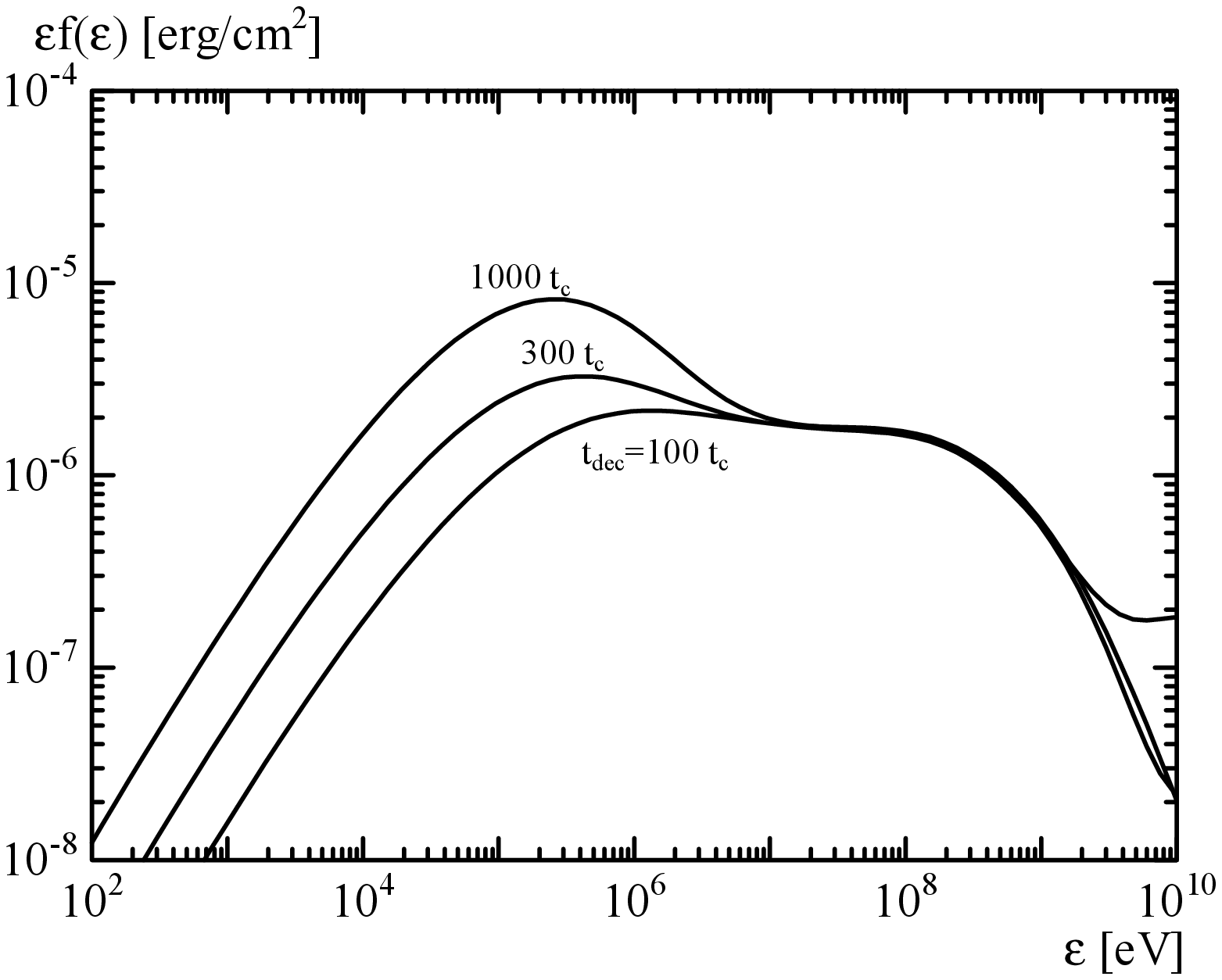}
\caption{
Spectra for the model of $D_{EE} \propto t^{-0.4}$,
$\gamma_{\rm inj}=10 \gamma_{\rm typ}$,
and different $t_{\rm dec}$ (100, 300, and 1000 $t_{\rm c,typ}$).
\label{fig:Tdep2}}
\end{figure}

We can see that the spectra above $\varepsilon_{\rm p}$
for $\chi=0.3$ and $0.4$ are well approximated by power-law functions.
Even for $\chi=0.5$, the spectrum from $\sim 500$ keV
to $\sim 3$ MeV can be accepted as a power-law function.
The peak energy $\varepsilon_{\rm p}$ for $\chi=0.3$ becomes above MeV,
because $t_{\rm acc}$ at $t=t_{\rm dec}$ is still shorter than
the final acceleration timescale assumed in advance, $\overline{\xi} t_{\rm c,typ}$.
For $\chi=0.5$, after $t=10 t_{\rm c,typ} \ll t_{\rm dec}=300 t_{\rm c,typ}$,
the timescale $t_{\rm coll}$ attains the upper limit $\overline{\xi} t_{\rm c,typ}$,
which makes $\gamma_{\rm e} \sim \gamma_{\rm typ}$.
Compared to the timescale of stay around $\gamma_{\rm inj}$,
the longer stayover around $\gamma_{\rm typ}$
yields the spectrum bump around a few hundred keV as set in advance.
So one may easily understand that the spectra in this model depend on
the timescale $t_{\rm dec}$.
Figure \ref{fig:Tdep2} shows that the spectral bump around
$\varepsilon_{\rm p}$ grows as $t_{\rm dec}$ extends.
It is apparent that the upper limit for $t_{\rm coll}$, whom we set up
to adjust $\varepsilon_{\rm p}$, causes the bumps.
If $t_{\rm coll}$ is elongated monotonically,
$\varepsilon_{\rm p}$ is determined by $t_{\rm coll}$ at $t=t_{\rm dec}$
as the case of $\chi=0.3$ in Figure \ref{fig:Tdep}.

The models of $D_{EE}$ in this section are toy models to demonstrate
the capability of MeV-GeV power-law spectrum in the slow heating model.
In addition, we have assumed the monochromatic injection of electrons.
The actual plasma turbulence and electron injection mechanism
may be more complicated than the models we tested.
The strength of magnetic fields may evolve with $D_{\rm EE}$,
or the first-order Fermi acceleration or surfing/drift/wake-field acceleration
\citep[see, e.g.,][]{ama07,hos08}
may work as the injection mechanism at shock front.
Although we need the nontrivial shape of $D_{EE}$ to reproduce
the high-energy power-law spectra,
the actual GRB plasmas may provide favorable conditions for MeV-GeV emissions.
PIC simulations can be strong tools to verify this scenario.
For example, \citet{cha08} show that the energy density of magnetic turbulence
in electron-positron plasma evolves as $\propto t^{-2/3}$ initially,
then steepens to $\propto t^{-1}$ later.
Such results encourage the model we discussed in this section.

\section{Delayed Emission}
\label{sec:del}

\begin{figure}[h]
\centering
\epsscale{1.0}
\plotone{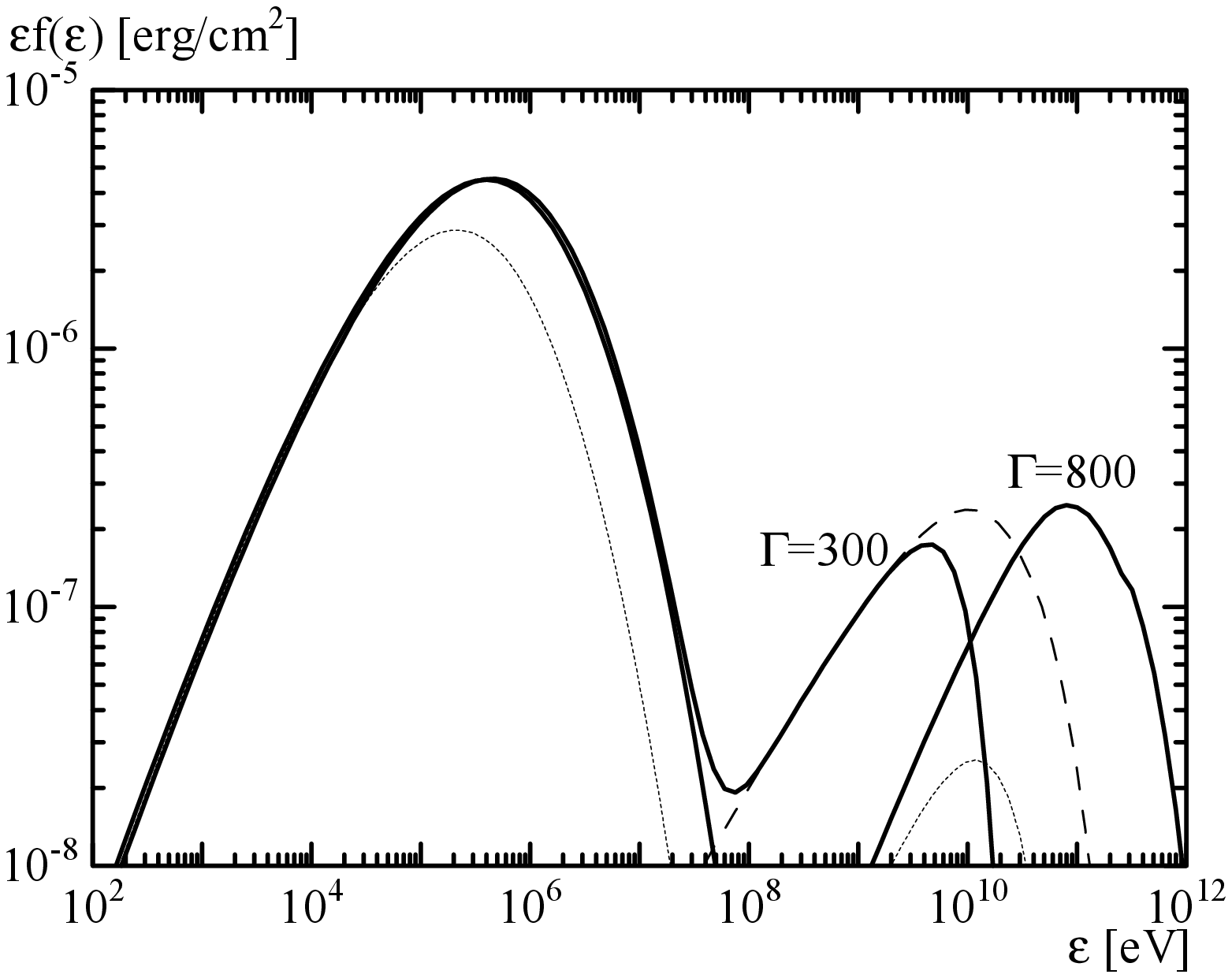}
\caption{
Spectra for the model of $B \propto e^{-t/t_{\rm dec}}$,
$\gamma_{\rm inj}=\gamma_{\rm typ}/10$,
$t_{\rm dec}=30 t_{\rm c,typ}$ and $t_{\rm sim}=t_{\rm dyn}$,
and different $\Gamma$ (solid lines, see the text).
The thin dotted line is the same as the spectrum in Figure \ref{fig:2nd},
where $t_{\rm sim}=t_{\rm dec} \ll t_{\rm dyn}$ and $B$ is constant.
The long dashed line is the IC component for $\Gamma=300$
without $\gamma \gamma$ absorption effect.
\label{fig:Gdel}}
\end{figure}

In the simulations discussed in the previous sections,
we have artificially halted the calculations at $t=t_{\rm dec}$
with constant magnetic fields.
However, the actual magnetic fields may not disappear suddenly,
and we may expect residual magnetic fields at $t>t_{\rm dec}$.
In this section, we consider emissions after the decay of magnetic fields.
Here, we assume a simple exponential decay and residual magnetic fields as
\begin{eqnarray}
B=\max\left( B_0 e^{-t/t_{\rm dec}},B_{\rm min} \right),
\end{eqnarray}
where $B_0$ and $B_{\rm min}$ are constants.
As for the acceleration timescale, to get rid of the heating effect
smoothly, we assume the rapid evolution of $t_{\rm coll}$ as
\begin{eqnarray}
t_{\rm coll}=\min\left(\bar{\xi} t_{\rm c,typ} e^{(t/t_{\rm dec})^2},t_{\rm dyn}\right).
\end{eqnarray}
We numerically follow electron cooling/heating and photon emission
during a period of $t_{\rm sim}=t_{\rm dyn}=l/c \gg t_{\rm dec}$
with $\gamma_{\rm inj}=\gamma_{\rm typ}/10$ and $t_{\rm dec}=30 t_{\rm c,typ}$
(see Figure \ref{fig:Gdel}).
Two parameter sets are adopted; one is the same as that in \S \ref{sec:simple}
($\Gamma=300$, $\gamma_{\rm typ}=3100$, $R=2.7 \times 10^{14}$ cm, $B_0=10^4$ G),
and another parameter set describes a higher $\Gamma$ case with
the same $\varepsilon_{\rm p}$ and $\Delta t$:
$\Gamma=800$, $\gamma_{\rm typ}=7800$, $R=1.9 \times 10^{15}$ cm, and $B_0=530$ G.
For the final magnetic fields, $B_{\rm min}=1$ G is adopted in both the two cases,
though there is no clue to the residual magnetic fields at present.
For $t \ll t_{\rm dec}$, the photon emission mechanism is the same as
those in \S \ref{sec:simple}.
However, after $t=t_{\rm dec}$, the main cooling mechanism is switched
from synchrotron to IC, because the photon density is assumed to be constant
within the shell of width $l$.


Comparing the thin dotted line with solid lines in Figure \ref{fig:Gdel},
it is clearly shown that the residual energy of electrons at $t=t_{\rm dec}$
is emitted via IC emission in GeV-TeV ranges.
The typical photon energy due to IC largely depends on $\Gamma$.
As shown by the long dashed line in Figure \ref{fig:Gdel},
$\gamma \gamma$ absorption affects the final spectrum for $\Gamma=300$
above 10 GeV (close to the cases of Figure \ref{fig:test}),
while it is negligible for $\Gamma=800$
owing to the lower photon density.
Therefore, this case indicates a delayed onset of GeV-TeV photons
compared to MeV photons with a timescale of $\sim t_{\rm dec}/\Gamma$.
On the other hand, the spectral shape in the low-energy region
is not altered in this model as seen in Figure \ref{fig:Gdel}.

Recent GRBs detected by {\it Fermi}-LAT tend to show
such a delayed onset of high-energy ($>100$ MeV) emission
\citep[GRB 080916C, GRB 080825C, etc.;][]{abd09,abd09b}.
However, it is noted that the broadband spectral shape
for GRB 080916C is not well reproduced by our simple toy model.
Future observations with {\it Fermi} or Cerenkov telescopes
will testify the model prospects and yield a clue to
the refinement of the model.

\begin{figure}[h]
\centering
\epsscale{1.0}
\plotone{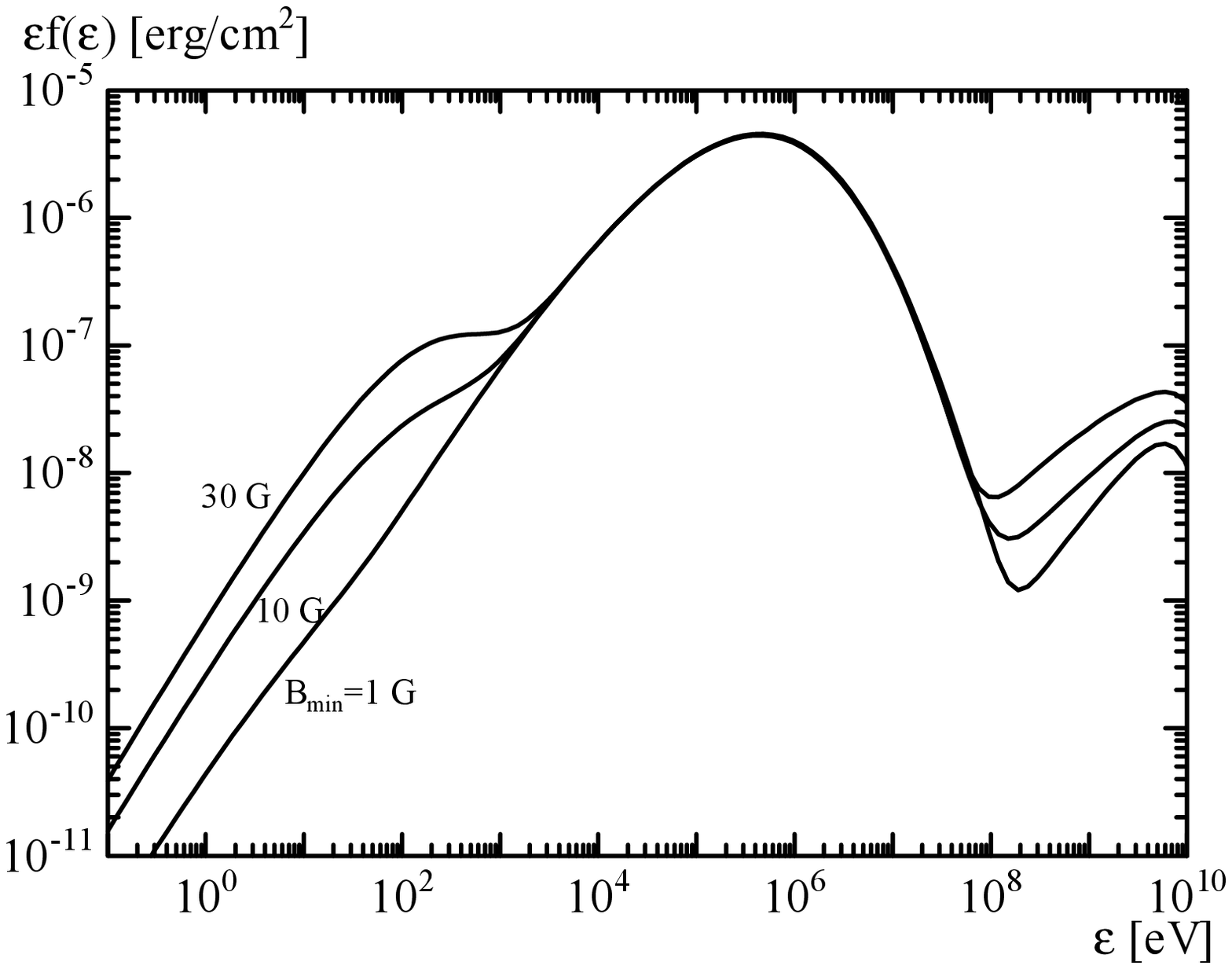}
\caption{
Spectra for the model of $B \propto e^{-t/t_{\rm dec}}$,
$\gamma_{\rm inj}=\gamma_{\rm typ}/10$,
$t_{\rm dec}=30 t_{\rm c,typ}$ and $t_{\rm sim}=10 t_{\rm dyn}$,
and different $B_{\rm min}$.
\label{fig:Odel}}
\end{figure}

Next, let us consider photon emission in longer timescales
than the dynamical timescale $t_{\rm dyn}$.
In the standard scenario, emission after the dynamical timescale may
be negligible because of the fast cooling of electrons.
However, in our scenario,
the residual electron energy may be released via synchrotron radiation,
which can contribute to optical emissions as seen in some GRBs
\citep{ves05,bla05}. 
Our results for such cases are shown in Figure \ref{fig:Odel},
where the parameters are the same as those for $\Gamma=300$ in Figure \ref{fig:Gdel},
but $t_{\rm sim}=10 t_{\rm dyn}$, $B_{\rm min}=1$, 10 and 30 G,
and rapid decay of the photon density $\propto \exp{[-(t/t_{\rm dec})^2]}$.
For simplicity, we neglect the effects of adiabatic cooling due to
shell expansion.

As seen in Figure \ref{fig:Odel}, optical synchrotron
emissions become more luminous
than the fluence of the power-law extrapolation from the X-ray spectra.
Most of the optical photons are emitted after $t=t_{\rm dyn}$
in the weak magnetic fields
so that broader pulse profile is expected for optical than
$\gamma$/X-ray bands.
Such longer variability timescales in prompt optical emission
is seen in GRB 080319B \citep{rac08}, though the flux
is brighter than the extrapolation from the X-ray spectra
by 3-4 orders of magnitude (1-2 orders in our results).
More luminous optical emissions as compared to X-ray
may be possible, if we change $t_{\rm dec}$.
Since our objective in this paper is not to reproduce spectra of specific GRBs,
we do not further discuss the fraction of optical flux here.

\section{Summary and Discussion}
\label{sec:sum}

Motivated by the energy transfer and low-energy spectrum problems,
we propose a new model to reproduce GRB prompt emissions.
In this model, electrons are continuously heated via plasma turbulences
within a timescale longer than the cooling timescale.
The acceleration timescale is assumed to be much longer than that
in the standard picture.
Emissions from cooled electrons are suppressed so that the low-energy
spectral index $\alpha$ is close to the observed value $-1$.
At least below MeV, the model spectrum does not contradict
the Band function very much.
Considering that most of the GRB spectra were obtained in energy ranges below MeV,
the model spectra may be consistent with a large fraction of GRBs.
In order to explain power-law spectra in MeV-GeV range observed in some GRBs,
we need a superposition of multiple components with different $\varepsilon_{\rm p}$,
or nontrivial shape and evolution of the diffusion coefficient.
Our model, under certain conditions, predicts delayed GeV-TeV emission via IC or
delayed optical emission with broad pulse profile
via synchrotron.
We expect that the accumulation of many GRB observations
will verify the characteristics predicted by our model in near future.

Roughly speaking, the energy release in this model is estimated
as $\Gamma N (t_{\rm dec}/t_{\rm c,typ})
\gamma_{\rm typ} m_{\rm e} c^2$, where $N$ is the total
number of accelerated electrons.
Given the total isotropic energy $E_{\rm iso}=4 \pi R^3 U$,
the required number of electrons depends on $t_{\rm dec}$.
Thus, the fraction of accelerated electrons can be much less than unity in this model,
while many authors have frequently assumed
that all electrons are accelerated \citep[see][]{eic05}.

One interesting point in our model is that the physical explanation
for the spectral peak energy $\varepsilon_{\rm p}$ is clear.
The balance between synchrotron cooling and heating provides us
the typical electron energy $\gamma_{\rm typ} m_{\rm e} c^2$,
from which we can estimate $\varepsilon_{\rm p}$.
The spectral peak and low-energy index are
reproduced by this mechanism unless $t_{\rm acc} \gg t_{\rm c}$.
One remaining problem is why shocked plasma in GRBs always adjust $t_{\rm coll}$
to make $\varepsilon_{\rm p}$ observed range of $\sim 1$ MeV.
To explain this we need another assumption; for example,
$D_{\rm EE} \propto U_B$, which implies
$t_{\rm acc} \propto B^{-2}$.
The balance between cooling time $t_{\rm c}
\propto B^{-2}$ and $t_{\rm acc}$ gives us $\gamma_{\rm typ} \propto B^0$.
Since we may write the luminosity as
$L_{\rm iso}=E_{\rm iso}/(R/c \Gamma^2)= 4 \pi (\epsilon_{\rm e}/\epsilon_B)
U_B R^2 \Gamma^2$, we obtain
\begin{eqnarray}
\varepsilon_{\rm p} \propto \Gamma B \gamma_{\rm typ}^2  \propto 
(\epsilon_{\rm e}/\epsilon_B)^{-1/2} L_{\rm iso}^{1/2} R^{-1},
\end{eqnarray}
where $\Gamma$-dependence disappears.
These results are consistent with the Yonetoku relation
\citep[$\varepsilon_{\rm p} \propto L_{\rm iso}^{0.5}$;][]{yon04},
except for the factor $R^{-1}$.
Of course, the above assumption is not trivial.
When we assume $D_{EE} \propto U_B^x$, the correlation becomes
$\varepsilon_{\rm p}
\propto 
L_{\rm iso}^{(3-2x)/2} \Gamma^{2x-2} R^{2x-3}$.
If the jitter radiation is applicable, in which the typical photon energy
depends on the coherent scale of the field,
the Yonetoku relation implies some correlations between
the typical scale of turbulence and luminosity.
In any case, fundamental studies of the long-term evolution of relativistic plasmas,
based on PIC simulations, etc.,
are indispensable to verify the model and luminosity correlations.

For the {\it Fermi}-LAT GRB 080916C, by a process of elimination,
\citet{zha09} conclude
that this GRB is emitted from a magnetically dominated outflow.
In such a case, the dissipation of the bulk kinetic energy
may not be due to internal shocks, because the Alfv\'en velocity
is very close to the light speed.
In order to produce non-thermal particles, the dissipation of
magnetic fields such as magnetic reconnection, etc. should occur
in the outflow.
The dissipation processes of magnetic fields may generate
magnetic turbulences so that we can expect
the second-order Fermi acceleration in this case too.



\acknowledgements

We appreciate the anonymous referee for the useful advice.
This work is partially supported by the Grant-in-Aid for
Scientific Research, No. 21540259 from the MEXT of Japan.

\end{document}